\documentclass[%
reprint,
superscriptaddress,
amsmath,amssymb,
aps,
pra,
]{revtex4-2}

\usepackage{graphicx}
\usepackage{dcolumn}
\usepackage{bm}
\graphicspath{{./figures/}}

\begin{document}

\title{Programmable Rapid Adiabatic Passage laser pulses \\for Ultra-fast Gates on trapped ions}

\author{En-Teng An}
\affiliation{Laboratory of Quantum Information, University of Science and Technology
of China, Hefei 230026, China }
\affiliation{Anhui Province Key Laboratory of Quantum Network, University of Science
and Technology of China, Hefei 230026, China}
\affiliation{ CAS Center For Excellence in Quantum Information and Quantum Physics,
University of Science and Technology of China, Hefei 230026, China }

\author{Hao-Qing Zhang}
\affiliation{Department of Physics, University of Colorado, Boulder, Colorado 80309, USA}

\author{Yun-Feng Huang}
\author{Chuan-Feng Li}
\author{Jin-Ming Cui}
\email[Corresponding author: ]{jmcui@ustc.edu.cn}
\affiliation{Laboratory of Quantum Information, University of Science and Technology
of China, Hefei 230026, China }
\affiliation{Anhui Province Key Laboratory of Quantum Network, University of Science
and Technology of China, Hefei 230026, China}
\affiliation{ CAS Center For Excellence in Quantum Information and Quantum Physics,
University of Science and Technology of China, Hefei 230026, China }
\affiliation{ Hefei National Laboratory, University of Science and Technology of China, Hefei 230088, China }

\date{\today}

\begin{abstract}
Scalable quantum gates remain a central challenge for trapped-ion quantum computing. Ultrafast gates driven by spin-dependent kicks (SDKs) provide a promising approach. However, current protocols rely on mode-locked lasers, suffering from inflexible timing control and limited single-SDK fidelity. To overcome this, we propose a scheme using rapid adiabatic passage (RAP) pulses modulated from a continuous-wave laser. We demonstrate that this RAP-based approach suppresses the sensitivity of SDKs to fluctuations in optical intensity, thereby enabling the construction of robust entangling gates. Furthermore, the programmable nature of these modulated pulses allows for precise control over pulse sequences, further optimizing gate performance.
\end{abstract}

\maketitle


\section{INTRODUCTION}
\begin{figure*}[t]
    \centering
    \includegraphics[scale=1]{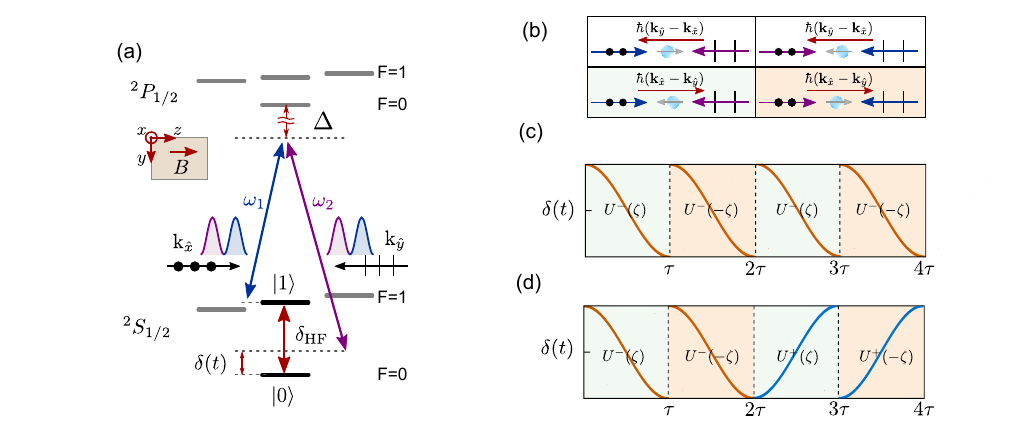}
    \caption{\label{fig:SDK_RAP} SDKs via RAP. (a) $^{171}\mathrm{Yb}^+$ energy levels and the RAP scheme. Qubits ($\left|0\right\rangle$, $\left|1\right\rangle$) are encoded in the hyperfine ground manifold. Two-photon Raman transitions transfer a spin-dependent momentum $\pm\hbar (\mathbf{k}_{\hat{x}} - \mathbf{k}_{\hat{y}})$, where $\hat{x}$ and $\hat{y}$ denote the polarization directions. The Raman beams, featuring a large single-photon detuning $\Delta$ and a time-dependent two-photon detuning $\delta(t)$, couple the Zeeman-split levels via $H$ and $V$ polarizations. (b) Spin-dependent momentum transfer. Counter-propagating Raman pulses aligned along the ion chain drive the axial phonons. The direction of the momentum kick depends on the initial spin state and the pulse propagation direction. (c) RAP sequence. A subsequent RAP pulse with an identical detuning sweep but opposite propagation direction cancels the accumulated dynamic phase, producing a pure momentum kick without spin-flip. (d) Error-mitigated RAP sequence. To compensate for incomplete dynamic phase cancellation arising from RAP pulse intensity mismatches, a second pair of RAP pulses with reversed detuning sweeps is appended.}
\end{figure*}
Trapped ion systems are among the most promising platforms for quantum computing, with exceptional coherence properties~\cite{monroe2013scaling}, near-perfect measurement fidelity~\cite{myerson2008high}, and high-fidelity entangling gates with minimal crosstalk~\cite{piltz2014trapped}. Two-qubit entangling gates typically employ the Mølmer-Sørensen (MS) scheme~\cite{milburn2000ion,sorensen2000entanglement}. Despite reaching high fidelities~\cite{ballance2016high}, their speed is fundamentally limited by the Lamb-Dicke regime, rendering them too slow for large-scale quantum computation~\cite{steane2000speed,leibfried2003quantum}.

Ultrafast gates based on spin-dependent kicks (SDKs) offer a promising route to accelerate gate operations~\cite{jaffe2018efficient,mizrahi2013ultrafast}. In this approach, ultrafast pulses entangle the ions' internal qubit states with their motional modes beyond the Lamb-Dicke regime~\cite{johnson2015sensing}. Precisely designed sequences drive these modes along closed phase-space trajectories, decoupling them from the internal states to yield pure qubit entanglement via the accumulated geometric phase~\cite{duan2004scaling,mizrahi2013ultrafast}. Consequently, realizing this method rigorously demands both high-fidelity SDKs and exact control over pulse timing.

Current ultrafast gate implementations rely on mode-locked lasers~\cite{wong2017demonstration,guo2022picosecond,torrontegui2020ultra,campbell2010ultrafast}, which suffer from two primary limitations. First, the pulse interval is fixed by the repetition rate, yielding sparse pulse sequences that limit optimization flexibility and degrade gate fidelity. For instance, earlier work employing a pulsed laser with an 86 MHz repetition rate required complex optical path adjustments for timing control, achieving a gate fidelity of only 76\%~\cite{wong2017demonstration}. While some schemes attempt to increase the repetition rate through resonant cavities and related techniques ~\cite{heinrich2019ultrafast,hussain2016ultrafast,hussain2021ultraviolet}, they introduce additional complexity that compromises stability. Second, conventional SDK implementations based on resonant Raman transitions~\cite{wong2017demonstration,shimizu2021ultrafast} are sensitive to variations in pulse power. The stringent requirement for long-term laser stability is experimentally demanding, ultimately limiting the single-SDK fidelity.

To overcome these challenges, we propose a programmable pulse scheme for robust ultrafast entangling gates. By modulating a continuous-wave (CW) laser, we implement SDKs based on rapid adiabatic passage (RAP), which significantly improves the robustness against laser intensity noise. Simulation results demonstrate that, for a single-pulse duration of 1 ns, the RAP-based SDK is highly insensitive to intensity noise in the sub-MHz regime, in stark contrast to conventional resonant Rabi SDK. Furthermore, by designing an alternating frequency-swept RAP sequence, we can further increase the upper bound of the tolerable noise frequency. Crucially, modulating the CW laser with an arbitrary waveform generator (AWG) enables arbitrary pulse sequences, completely avoiding the constraints of fixed repetition rates. Simulations show that for a given single-SDK fidelity, the gate fidelity of the programmable pulses significantly outperforms that of conventional mode-locked pulses (which typically have repetition rates in the tens of MHz). To match this performance, the mode-locked laser would require a repetition rate on the order of hundreds of MHz. Moreover, as the target single-SDK fidelity increases , this repetition rate requirement will become even more stringent.
\section{Spin-dependent kick in the $^{171}\mathrm{Yb}^+$ system}

An SDK imparts a state-dependent momentum to a trapped ion by driving a stimulated Raman transition with a pair of counter-propagating pulses. During this process, the ion absorbs a photon from one beam and undergoes stimulated emission into the other, thereby acquiring a net momentum of $\pm 2\hbar k$ depending on its internal state ($\left|0\right\rangle$ or $\left|1\right\rangle$). We analyze this SDK mechanism using the full eight-level hyperfine structure of $^{171}\mathrm{Yb}^{+}$.

We model the eight energy levels of the $^{171}\mathrm{Yb}^{+}$ ion, as illustrated in Fig.~\ref{fig:SDK_RAP}(a). Qubits are encoded in the ground hyperfine states: $^{2}S_{1/2}\left|F=0,m_{F}=0\right\rangle \equiv |0\rangle$ and $^{2}S_{1/2}\left|F=1,m_{F}=0\right\rangle\equiv |1\rangle$, with an energy splitting $\delta_{\mathrm{HF}}=2\pi\times12.6428$~GHz. The Raman beams are detuned by hundreds of GHz from the $^{2}P_{1/2}$ manifold, a detuning substantially smaller than the 100~THz splitting between $^{2}P_{1/2}$ and $^{2}P_{3/2}$. We therefore neglect coupling to the $^{2}P_{3/2}$ manifold and treat $^{2}P_{1/2}$ as the virtual energy level. The hyperfine splitting in $^{2}P_{1/2}$ is $\delta_{1}=2\pi\times2105$~MHz. We account for the linear Zeeman effect ($2\pi\times1.4$~MHz/G for $^{2}S_{1/2}$ and $2\pi\times0.47$~MHz/G for $^{2}P_{1/2}$) and quadratic Zeeman effect ($2\pi\times310.8~\mathrm{Hz}/\mathrm{G}^{2}$) under a DC magnetic field that defines the quantization axis. 

The qubit states $\left|0\right\rangle$ and $\left|1\right\rangle$ exhibit long coherence times~\cite{lu2023measurement} since their transition constitutes a clock transition. This transition is driven via $|L\rangle$ or $|R\rangle$ polarization. We employ linearly polarized light ($|H\rangle$ and $|V\rangle$) to drive the Raman process, decomposed into circular polarizations as:  
\begin{equation}  
    \begin{aligned}  
        |H\rangle &= \frac{1}{\sqrt{2}}\left(|L\rangle+|R\rangle\right)\\  
        |V\rangle &= \frac{1}{\sqrt{2}}\left(|L\rangle-|R\rangle\right).  
    \end{aligned}  
\end{equation}  

The effective two-level Hamiltonian is given by (see Appendix~\ref{app:effective_hamiltonian} for a detailed derivation)
\begin{equation}
    H_{\mathrm{I}} = \frac{\hbar}{2}\left[\Omega(t)\left(e^{i2\zeta kx}\sigma_+ + e^{-i2\zeta kx}\sigma_-\right) + \delta(t)\sigma_z\right],
    \label{eq:hamiltonian}
\end{equation}
where $\Omega(t)$ is the effective Rabi frequency (which is proportional to $\sqrt{I_{\hat{x}}I_{\hat{y}}}/\Delta$), $x$ denotes the position of the ion, and $\delta(t)$ represents the time-dependent two-photon detuning. The effective wave-vector difference between the two pulses along the motional axis is $\Delta k = |\mathbf{k}_{\hat{x}}-\mathbf{k}_{\hat{y}}|  \approx 2\zeta k$, with $\zeta = \pm 1$ depending on the incidence direction of the pulses.

In a standard SDK based on resonant Raman transitions, the two-photon detuning $\delta(t)$ is set to zero. A complete qubit flip is achieved when the pulse area satisfies $\int_0^\tau \Omega(t)dt = \pi$, where $\tau$ is the pulse duration. Because the pulse duration is much shorter than the ions' motional period, the free evolution of the motion can be neglected. The unitary evolution operator for a single Raman pulse pair then simplifies to
\begin{equation}
    U_{\pi}(\zeta) = e^{i2\zeta kx}\sigma_+ + \text{H.c}.
    \label{eq:unitary_x}
\end{equation}

The ion position is expressed in terms of the motional mode: $x = x_0(a + a^\dagger)$, where $x_0 = \sqrt{\hbar/(2M\omega)}$ is the zero-point spread of the mode with ion mass $M$ and mode frequency $\omega$. By defining the Lamb-Dicke parameter as $\eta = 2k x_0$, the unitary operator is rewritten as:
\begin{equation}
    U_{\pi}(\zeta) = \hat{D}(i \zeta \eta\sigma_z),
    \label{eq:unitary_D}
\end{equation}
where $\hat{D}(\alpha) = \exp(\alpha a^\dagger - \alpha^* a)$ is the displacement operator for the motional mode. This equation shows that for a fixed $\Delta k$ direction, the ion undergoes an instantaneous phase-space displacement, with the displacement direction conditioned strictly on its internal spin state, as shown in Fig.~\ref{fig:SDK_RAP}(b).

\section{ Rapid Adiabatic Passage Pulses}
\label{3}
\subsection{Single RAP evolution}
RAP is an established technique for robust quantum state control through adiabatic parameter modulation~\cite{PhysRevA.45.5297}. By slowly sweeping the detuning of a driving field through resonance, RAP maintains the system in an instantaneous eigenstate. Compared to conventional resonant $\pi$ pulses, RAP is insensitive to variations in laser frequency, pulse amplitude, and timing. This inherent robustness suits it for inhomogeneously broadened systems~\cite{vitanov2001laser} and enables reliable population inversion across many physical platforms~\cite{drain1949direct,slichter1961adiabatic,janzen1968adiabatic}. Here we integrate the RAP technique into the SDK sequence.

The RAP pulse profiles are defined by time-dependent intensity envelopes $I_{\hat{x},\hat{y}}(t)$ and detuning $\delta(t)$
\begin{equation}
    \begin{aligned}
        I_{\hat{x}}(t) &= I_{\hat{y}}(t) = I_0\sin^2(\pi t/\tau),\\
        \delta(t) &= \delta_0\cos(\pi t/\tau),
    \end{aligned}
    \label{eq:laser_profile}
\end{equation}
where $\tau$ denotes the total duration of a single spin-flip operation. Experimentally, these pulse profiles can be synthesized by modulating a CW laser. The intensity envelopes are shaped using a Mach-Zehnder electro-optic modulator (MZ-EOM), while the time-dependent detuning is controlled by an electro-optic phase modulator. Driven by a synchronized high-speed AWG, both modulators enable arbitrary programming of the laser for precise generation of the desired RAP waveforms. 

\begin{figure}[t]
    \centering
    \includegraphics[scale=1]{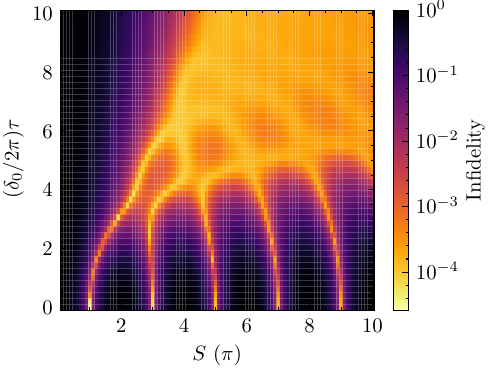}
    \caption{\label{RAP_infidelity}Infidelity of the RAP population transfer under various parameters. The fidelity is evaluated as a function of the frequency sweep amplitude $\delta_0$ and the pulse area $S = \int_0^\tau \Omega(t) \, dt$. The single-photon detuning is set to $\Delta = 2\pi \times 800$~GHz, and the duration of a single RAP pulse is fixed at $\tau = 1$~ns.}
\end{figure}
\begin{figure*}[t]
    \centering
    \includegraphics[scale=1]{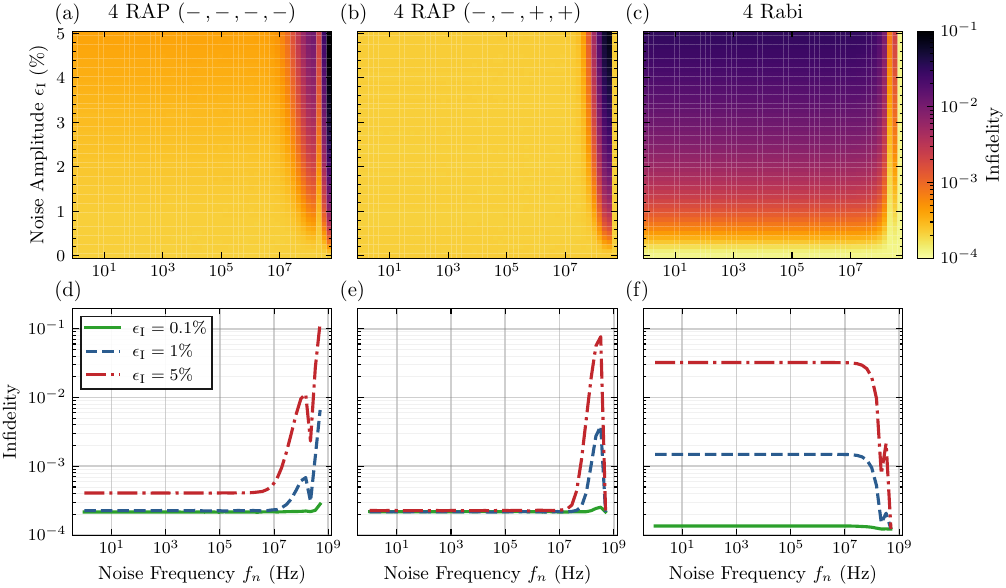}
    \caption{\label{robustness} Simulated infidelity of the four-pulse SDK sequences under intensity noise. (a)-(c) Two-dimensional colormaps illustrating the infidelity as a function of noise frequency $f_{\mathrm{n}}$ and noise amplitude $\epsilon_{\mathrm{I}}$ for the $\mathrm{RAP}_{\mathrm{uni}}$ sequence [labeled as 4 RAP ($-,-,-,-$)], the $\mathrm{RAP}_{\mathrm{alt}}$ sequence [4 RAP ($-,-,+,+$)], and the traditional 4 Rabi sequence, respectively. (d)-(f) Infidelity as a function of noise frequency $f_{\mathrm{n}}$ evaluated at fixed noise amplitudes of $0.1\%$, $1\%$, and $5\%$, corresponding to the sequences in (a)-(c). The pulse interval is set to $\tau = 1$~ns. The traditional Rabi-SDK utilizes standard $\pi$ pulses, whereas the RAP-SDK sequences employ the optimized parameters $\delta_0 = 9\pi/\tau$ and $S = 4.02\pi$.}
\end{figure*}
Driven by the RAP pulse, the system evolves adiabatically along the instantaneous eigenstate of the time-dependent Hamiltonian, resulting in a robust spin flip at the end of the pulse. During this process, the system accumulates a dynamical phase depending on both the frequency sweep direction and the initial state of the system. For a negative sweep (i.e., the detuning $\delta(t)$ sweeps from positive to negative), the effective adiabatic evolution operator is given by (see Appendix~\ref{app:evolution_RAP} for a detailed derivation):
\begin{equation}
        U^-(\zeta) = -e^{i(2\zeta kx+\Theta(\tau))}\sigma_+ + e^{-i(2\zeta kx+\Theta(\tau))}\sigma_-,
\end{equation}
where $\Theta(\tau) = \frac{1}{2}\int_0^\tau \sqrt{\Omega(t)^2 + \delta(t)^2} dt$ is the accumulated dynamic phase. 
\subsection{Robust cancellation of the dynamic phase}

To eliminate this dynamic phase and achieve a pure SDK, the simplest approach is to apply a second RAP pulse with an identical detuning sweep but an opposite propagation direction, as shown in Fig.~\ref{fig:SDK_RAP}(c). Ideally, due to the ultrashort duration of the RAP operations, the laser intensity is minimally affected by temporal fluctuations. This allows the accumulated dynamic phases of two consecutive RAP operations to be approximately identical. Because both pulses share the same sweep, the ion follows different instantaneous eigenstate branches in each pulse, resulting in dynamic phases with opposite signs that effectively cancel. As a result of this phase cancellation, the two-pulse sequence yields a purely state-dependent momentum kick of $4\hbar k$ without flipping the internal spin state:
\begin{equation}
    U^{(2)}_{\mathrm{SDK}} = U^-(-\zeta)U^-(\zeta) = \exp\left(-i4\zeta kx\sigma_z\right).
    \label{eq:pure_SDK}
\end{equation}

To further mitigate the residual dynamic phase errors caused by minimal temporal fluctuations in the laser intensity, we implement a composite SDK sequence comprising four RAP pulses, following the approach of Qiu \textit{et al.}~\cite{qiu2022spinor}, as illustrated in Fig.~\ref{fig:SDK_RAP}(d). By symmetrically arranging the pulses, this four-pulse sequence effectively eliminates dynamic phase errors arising from both static offsets and linear temporal drifts in the laser intensity.

To explicitly demonstrate this error cancellation, we consider a time-dependent perturbation to the Rabi frequency,
\begin{equation}
    \Omega(t) = \Omega_0(t)[1+\epsilon_\Omega(t)].
\end{equation}
Substituting this perturbed Rabi frequency into the definition of the dynamic phase and keeping only the leading-order term with respect to the perturbation amplitude yields
\begin{equation}
    \Delta \Theta \approx \int_0^{\tau_p} \epsilon_\Omega(t) \frac{\Omega_0(t)^2}{\sqrt{\Omega_0(t)^2 + \delta(t)^2}} \,dt,
\end{equation}
where $\tau_p$ is the duration of a single RAP pulse. Assuming the intensity fluctuation $\epsilon_\Omega(t)$ varies slowly enough to be treated as a constant during a single pulse operation, we can extract it from the integral. By defining a sequence-specific constant $C = \int_0^{\tau_p} \frac{\Omega_0(t)^2}{\sqrt{\Omega_0(t)^2 + \delta(t)^2}} \,dt$, the phase error for a single pulse is approximately $\Delta \Theta \approx C \epsilon_\Omega(t)$. 

Consequently, for the four-pulse sequence shown in Fig.~\ref{fig:SDK_RAP}(d) with a pulse spacing of $\tau$, the total residual dynamic phase induced by intensity fluctuations can be summed as
\begin{equation}
    \Delta \Theta_{\mathrm{total}} \approx C \left[ \epsilon_\Omega(\tau) - \epsilon_\Omega(2\tau) - \epsilon_\Omega(3\tau) + \epsilon_\Omega(4\tau) \right].
\end{equation}
Expanding the time-dependent perturbation $\epsilon_\Omega(t)$ in a Taylor series with respect to time reveals that both the zeroth-order (constant offset) and first-order (linear drift) terms cancel out exactly. This symmetric cancellation leaves only the second-order residual phase,
\begin{equation}
    \Delta \Theta_{\mathrm{total}} \approx 2 C \epsilon_\Omega''(0) \tau^2 + \mathcal{O}(\tau^3).
\end{equation}
Because these leading-order phase errors are robustly eliminated by the sequence design, the residual phase accumulation becomes negligible, allowing the actual evolution to closely approximate the ideal dynamics. Consequently, the effective evolution operator for this four-pulse SDK sequence effectively reduces to a purely state-dependent momentum kick of $8\hbar k$:
\begin{equation}
    \begin{aligned}
        U^{(4)}_{\mathrm{SDK}} &= U^+(-\zeta)U^+(\zeta)U^-(-\zeta)U^-(\zeta) \\
        &\approx \exp\left(-i8\zeta kx\sigma_z\right).
    \end{aligned}    
\end{equation}

\subsection{Fidelity analysis}
To demonstrate the robustness of the RAP-based SDK we analyze the impact of laser intensity noise at different frequencies on the SDK fidelity. The noise model for the optical intensity fluctuations is given by
\begin{equation}
I_{\mathrm{actual}}(t) = I_0\sin^2(\pi t/\tau)\left(1 + \epsilon_{\mathrm{I}}\sin\left(2\pi f_{\mathrm{n}} t_{\mathrm{n}} + \phi_{\mathrm{n}}\right)\right),
\end{equation}
where $f_{\mathrm{n}}$ is the noise frequency, $t_{\mathrm{n}} \in \{\tau/2, 3\tau/2, \dots\}$ is the center time of an individual RAP pulse, $\phi_{\mathrm{n}}$ is a random phase, and $\epsilon_{\mathrm{I}}$ is the amplitude of the intensity fluctuations. The ideal evolution operator of the SDK is defined as
\begin{equation}
    U(\phi_k) = \begin{pmatrix} e^{-i\phi_k} & 0 \\ 0 & e^{i\phi_k} \end{pmatrix},
\end{equation}
where $\phi_k$ is the phase induced by the wavevector difference, which is essential for generating the SDK. However, in practical experiments, due to factors such as spontaneous emission and laser intensity fluctuations, the actual SDK evolution operator deviates from the ideal case, leading to a degradation in the SDK fidelity. The fidelity is thus evaluated as
\begin{equation}
F_{\mathrm{SDK}} = \langle \left| \langle\psi_i|U^\dagger(\phi_k)\tilde{U}(\phi_k,\epsilon_{\mathrm{I}})|\psi_i\rangle \right|^2 \rangle_{\psi_i, \phi_{\mathrm{n}}},
\end{equation}
where $\tilde{U}(\phi_k,\epsilon_{\mathrm{I}})$ represents the actual evolution operator under experimental noise, and the outer brackets denote an ensemble average over all possible initial states $|\psi_i\rangle$ and the random noise phase $\phi_{\mathrm{n}}$. Here, $|\psi_i\rangle$ are chosen as the six eigenstates of the Pauli operators, which form a spherical 2-design on the Bloch sphere.

Fig.~\ref{RAP_infidelity} shows the infidelity of a single RAP population transfer as a function of the frequency sweep amplitude $\delta_0$ and the pulse area $S = \int_0^\tau \Omega(t) dt$. In our simulations, we consider a typical trapped-ion setup with a single-photon detuning of $\Delta = 2\pi \times 800$~GHz, where the duration of a single RAP pulse is fixed at $\tau = 1$~ns. The RAP population transfer exhibits significant robustness over a broad range of frequency sweep amplitudes and pulse areas. Based on these numerical results, we select a set of parameters ($\delta_0 = 9\pi/\tau$ and $S = 4.02\pi$) that provides high fidelity and robustness for subsequent analysis. Fig.~\ref{robustness} illustrates the dependence of the fidelity on the noise amplitude and frequency for the traditional four-pulse resonant Rabi SDK, as well as for two four-pulse RAP-SDK sequences with frequency sweep directions of $(-,-,-,-)$ and $(-, -, +, +)$, which we hereafter denote as $\mathrm{RAP}_{\mathrm{uni}}$ and $\mathrm{RAP}_{\mathrm{alt}}$, respectively. Because the noise model defined above has a periodicity of $1/(2\tau)$, we restrict our analysis to the noise frequency range of $[0, 1/(2\tau)]$, which corresponds to $[0, 500$~MHz].

As shown in Figs.~\ref{robustness}(a)-\ref{robustness}(c), both RAP-SDK sequences demonstrate superior robustness over the traditional Rabi-SDK at frequencies below 10 MHz. To systematically analyze the frequency dependence, we evaluate the response of the three pulse sequences to noise across the specified amplitudes of $0.1\%$, $1\%$, and $5\%$ [Figs.~\ref{robustness}(d)-\ref{robustness}(f)]. Below $10$~MHz, the fidelities of all sequences are largely frequency-independent. In the high-frequency regime, the sequences exhibit distinct compensation behaviors. For the Rabi SDK, fidelity increases sharply above $1/(8\tau)$ and becomes almost entirely immune to noise at $1/(2\tau)$. At this specific frequency, the intensity noise phases of adjacent resonant pulses are exactly opposite, compensating for each other to effectively form perfect $\pi$ pulses.

In contrast, the RAP-SDK sequences are dominated by the dynamic phase, which accumulates at higher frequencies and generally degrades fidelity. Notably, the $\mathrm{RAP}_{\mathrm{alt}}$ sequence exhibits a broader tolerance to high-frequency noise than the $\mathrm{RAP}_{\mathrm{uni}}$ sequence. Furthermore, specific geometric cancellations occur depending on the sweep pattern. At $f_{\mathrm{n}} = 1/(4\tau)$, the fidelity of the $\mathrm{RAP}_{\mathrm{alt}}$ sequence drops to a minimum, whereas the $\mathrm{RAP}_{\mathrm{uni}}$ sequence unexpectedly recovers. This happens because noise perturbations on every other RAP pulse (i.e., separated by $2\tau$) exhibit opposite phases, perfectly canceling their residual dynamic phases. At $f_{\mathrm{n}} = 1/(2\tau)$, the situation entirely reverses: the $\mathrm{RAP}_{\mathrm{alt}}$ sequence regains high fidelity, while the $\mathrm{RAP}_{\mathrm{uni}}$ sequence plunges to its minimum and becomes highly sensitive to noise.

In practical experiments, laser intensity fluctuations are predominantly concentrated in the low-frequency region, typically below the kHz range. As the frequency increases, the noise intensity decreases, exhibiting a characteristic $1/f$ trend \cite{Savard1997,Wang2020}. As demonstrated above, both RAP-SDK sequences exhibit exceptional robustness in this critical low-frequency regime compared to the traditional Rabi-SDK. Therefore, the proposed RAP-SDK sequences effectively decouple the quantum operations from the dominant experimental noise sources, offering a substantial practical advantage for achieving highly robust ultrafast gates.
\section{fast entangling gate with spin-dependent kicks}
The fast gate framework was introduced by Garc{\'i}a-Ripoll \textit{et al.}~\cite{garcia2003speed}. Its core principle applies a sequence of spin-dependent momentum kicks (SDKs) to induce transient state-dependent forces, allowing the system to accumulate an entangling geometric phase while ensuring that motional modes decouple from the internal states at the end of the operation. In this framework, optimizing the timing sequence of the SDK pulses is central to gate construction. We consider a two-ion system where laser pulses propagate along the ion chain to drive the axial motional modes. For simplicity, we utilize an SDK composed of two RAP pulses with identical frequency sweep directions. The two-ion SDK operator can be expressed as a product of single-ion operators:
\begin{equation}
    U(\zeta) = \exp\left[-i4\zeta k (x_1\sigma_z^1 + x_2\sigma_z^2)\right],
\end{equation}
where $x_{1,2}$ denote the position operators of the two ions, and $\sigma_z^{1,2}$ are the Pauli-z operators acting on the respective internal states. By transforming the position operators into the motional mode operators, we can rewrite the evolution operator in terms of displacement operators:
\begin{equation}
    U(\zeta) = \hat{D}_{\mathrm{c}}\left[-2i\zeta\eta_{\mathrm{c}}(\sigma_z^1 + \sigma_z^2)\right] \hat{D}_{\mathrm{s}}\left[-2i\zeta\eta_{\mathrm{s}}(\sigma_z^1 - \sigma_z^2)\right].
    \label{eq:SDK_two_ion}
\end{equation}
Here, the effective Lamb-Dicke parameters are defined as
\begin{equation}
    \eta_{\mathrm{c}} = k\sqrt{\frac{\hbar}{M\omega_{\mathrm{c}}}}, \quad \eta_{\mathrm{s}} = k\sqrt{\frac{\hbar}{M\omega_{\mathrm{s}}}},
\end{equation}
where $\omega_{\mathrm{c}}$ and $\omega_{\mathrm{s}}$ are the frequencies of the center-of-mass (COM) and stretch mode, respectively, satisfying $\omega_{\mathrm{s}} = \sqrt{3}\omega_{\mathrm{c}}$. Let $z_k = -n_k\zeta$, where $n_k$ denotes the number of consecutive kicks applied at time $t_k$. Assuming the complete gate operation consists of $N$ such kick groups, the pulse sequence is characterized by the following $2N$ parameters:
\begin{equation}
    \begin{aligned}
        \mathbf{z} &= \{z_1, z_2, \dots, z_N\}, \\
        \mathbf{t} &= \{t_1, t_2, \dots, t_N\}.
    \end{aligned}
    \label{eq:pulse_parameters}
\end{equation}
In the interaction picture, the total evolution operator for the gate operation can be expressed as (see Appendix~\ref{app:total_gate} for a detailed derivation)
\begin{equation}
    U_{\mathrm{total}} = \hat{D}_{\mathrm{c}}(\alpha_{\mathrm{c}}\Sigma_{\mathrm{c}})\hat{D}_{\mathrm{s}}(\alpha_{\mathrm{s}}\Sigma_{\mathrm{s}}) \exp(-i\Phi\sigma_z^1\sigma_z^2),
\end{equation}
where $\Sigma_{\mathrm{c}} = \sigma_z^1 + \sigma_z^2$ and $\Sigma_{\mathrm{s}} = \sigma_z^1 - \sigma_z^2$ denote the symmetric and antisymmetric spin operators, respectively. The motional displacements and the accumulated geometric phase are given by
\begin{equation}
    \begin{gathered}
        \alpha_{\mathrm{c}} = 2i\eta_{\mathrm{c}}\sum_{k=1}^N z_k e^{i\omega_{\mathrm{c}} t_k}, \\
        \alpha_{\mathrm{s}} = 2i\eta_{\mathrm{s}}\sum_{k=1}^N z_k e^{i\omega_{\mathrm{s}} t_k}, \\
        \Phi = \sum_{j=2}^N \sum_{k=1}^{j-1} 8 z_j z_k \left\{ \eta_{\mathrm{s}}^2 \sin[\omega_{\mathrm{s}}(t_j - t_k)] - \eta_{\mathrm{c}}^2 \sin[\omega_{\mathrm{c}}(t_j - t_k)] \right\}.
    \end{gathered}
    \label{eq:gate_conditions}
\end{equation}
The ideal two-qubit gate is given by $U_{\mathrm{ideal}} = \exp(-i\pi\sigma_z^1\sigma_z^2/4)$, which requires satisfying the conditions $\alpha_{\mathrm{c}} = 0$, $\alpha_{\mathrm{s}} = 0$, and $\Phi = \pi/4$. The gate fidelity can then be expressed as~\cite{bentley2015trapped}
\begin{equation}
    \begin{aligned}
        F_{\mathrm{o}} &= \frac{1}{12} \Big\{ 6 + e^{-8m_{\mathrm{c}}|\alpha_{\mathrm{c}}|^2} + e^{-8m_{\mathrm{s}}|\alpha_{\mathrm{s}}|^2} \\
        &\quad + 4 e^{-2(m_{\mathrm{c}}|\alpha_{\mathrm{c}}|^2 + m_{\mathrm{s}}|\alpha_{\mathrm{s}}|^2)} \cos[2(\Phi-\pi/4)] \Big\},
    \end{aligned}
    \label{eq:fidelity}
\end{equation}
where $m_{\mathrm{c},\mathrm{s}} = \bar{n}_{\mathrm{c},\mathrm{s}} + 1/2$, with $\bar{n}_{\mathrm{c}}$ and $\bar{n}_{\mathrm{s}}$ being the initial average phonon numbers of the COM and stretch modes, respectively.
\begin{figure}[t]
    \centering
    \includegraphics[scale = 1]{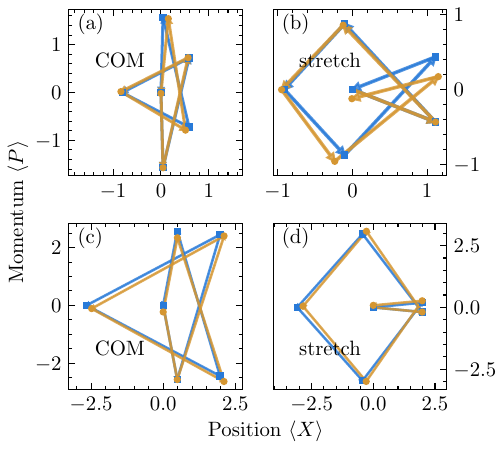}
    \caption{ \label{fig:space}Phase-space trajectories of the COM and stretch phonon modes in the rotating frame for the (a, b) GZC scheme and (c, d) FRAG scheme with $n=3$. (a, c) Trajectories of the COM mode. (b, d) Trajectories of the stretch mode. To decouple the spin and motional states, the trajectories must form closed loops at the end of the gate. The blue lines with squares represent the programmable pulse scheme, with perfect trajectory closure. The yellow lines with circles correspond to conventional pulsed lasers at a $50$~MHz repetition rate. Due to the discrete timing constraint, the conventional scheme shows residual displacements at the end of the gate, indicating imperfect closure and residual spin-motion entanglement.}
\end{figure}
The objective of pulse sequence optimization is to minimize the gate infidelity $1 - F_{\mathrm{o}}$. Given the $2N$ parameters involved, a brute-force search for the global minimum is often computationally intractable. Various schemes have been proposed to reduce the optimization complexity by constraining the parameter space~\cite{garcia2003speed,savill2025error,ratcliffe2020micromotion,gale2020optimized,bentley2015trapped,bentley2013fast}. Two classic examples are the Garc\'{i}a-Ripoll, Zoller, and Cirac (GZC) scheme~\cite{garcia2003speed} and the fast robust antisymmetric gate (FRAG) scheme~\cite{bentley2013fast}. The GZC scheme is defined by
\begin{equation}
    \begin{aligned}
        z &= (-2n, 3n, -2n, 2n, -3n, 2n),\\
        t &= (-\tau_1, -\tau_2, -\tau_3, \tau_1, \tau_2, \tau_3).
    \end{aligned}
\end{equation}
The FRAG scheme is characterized by
\begin{equation}
    \begin{aligned}
        z&= (-n, 2n, -2n, 2n, -2n, n),\\
        t&= (-\tau_1, -\tau_2, -\tau_3, \tau_1, \tau_2, \tau_3).
    \end{aligned}
\end{equation}
\begin{figure*}[t]
    \centering
    \includegraphics[scale=1]{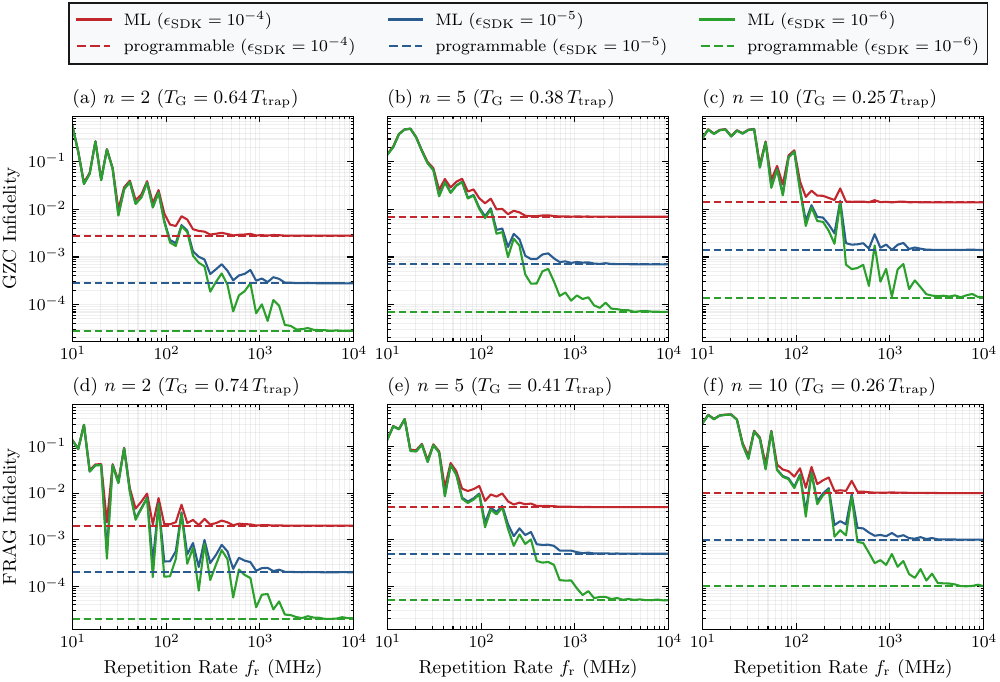}
    \caption{\label{fig:gate_infidelity} Simulated gate infidelity of the GZC and FRAG schemes as a function of the laser repetition rate $f_{\mathrm{r}}$. (a)-(c) Gate infidelity for the GZC scheme with the sequence parameter $n = 2, 5$, and $10$, respectively. (d)-(f) Corresponding gate infidelity for the FRAG scheme with $n = 2, 5$, and $10$. The solid curves represent the performance of conventional mode-locked pulses, where the infidelity generally decreases at higher repetition rates and asymptotically approaches the performance of the programmable pulses. In contrast, the horizontal dashed lines denote the performance of the proposed programmable pulse scheme. The distinct colors indicate different baseline single-SDK errors $\epsilon_{\mathrm{SDK}}$ (red: $\epsilon_{\mathrm{SDK}} = 10^{-4}$, blue: $\epsilon_{\mathrm{SDK}} = 10^{-5}$, green: $\epsilon_{\mathrm{SDK}} = 10^{-6}$). The total gate times $T_{\mathrm{G}}$, expressed in units of the trap period $T_{\mathrm{trap}}$, are labeled above each panel.}
\end{figure*}

For traditional mode-locked (ML) pulsed lasers, the pulse timing is fixed by the repetition rate, which limits the optimization of the pulse sequence. As illustrated in Fig.~\ref{fig:space}, this discrete timing constraint leads to imperfect closure of the phase-space trajectories at the end of the gate, leaving residual spin-motion entanglement. In contrast, our programmable pulse scheme permits arbitrary timing control, ensuring perfect trajectory closure so we can optimize the pulse sequence for better gate performance. To evaluate the two pulse schemes, we optimize the pulse sequences under both the GZC and FRAG frameworks. We compare the gate fidelity and total gate duration achieved using programmable CW-modulated pulses against conventional ML pulses across several repetition rates. Since the final gate fidelity depends on both the algorithmic optimization and the execution of individual SDKs, the total gate fidelity is defined as
\begin{equation}
    F_{\mathrm{G}} \approx (F_\mathrm{SDK})^{N_{\mathrm{p}}} F_{\mathrm{o}},
    \label{eq:total_fidelity}
\end{equation}
where $N_{\mathrm{p}}$ is the total number of SDKs required to construct the entangling gate and $F_{\mathrm{o}}$ is the algorithmic fidelity derived from timing optimization. $F_\mathrm{SDK}$ is the fidelity of a single SDK, and its infidelity is expressed as $\epsilon_{\mathrm{SDK}} = 1 - F_\mathrm{SDK}$. For simplicity, we assume that both motional modes are initially cooled to the ground state, i.e., $\bar{n}_{\mathrm{c},\mathrm{s}} = 0$.

Fig.~\ref{fig:gate_infidelity} illustrates the gate infidelities of the GZC and FRAG schemes under various single-SDK infidelities and gate times. The solid and dashed lines represent the infidelities of gates executed using ML pulses and programmable pulses, respectively, plotted as a function of the repetition rate. As the repetition rate increases, the fidelity achieved by the ML pulses asymptotically approaches that of their programmable counterparts. This behavior indicates that at high repetition rates, timing-optimization errors become negligible, leaving the accumulated SDK error as the dominant source of infidelity. 

Furthermore, shortening the gate time $T_{\mathrm{G}}$ noticeably flattens the decay curve of the ML infidelity. This slower convergence indicates that a correspondingly higher repetition rate is required to achieve the same algorithmic fidelity $F_{\mathrm{o}}$. However, this effect is counterbalanced by hardware constraints: since the gate time scales with the SDK number as $T_{\mathrm{G}} \propto N_{\mathrm{p}}^{-2/3}$, a shorter gate requires more SDKs, which inherently raises the accumulated error floor. Because both the timing-optimization errors and the accumulated SDK error floor scale concurrently as the gate time decreases, the relative threshold at which timing errors become overshadowed by SDK errors is preserved. Consequently, the critical repetition rate—at which operations driven by ML pulses match the fidelity of their programmable counterparts—remains independent of the gate time. 

Moreover, to fully exploit the benefits of a higher single-SDK fidelity, timing-optimization errors must be suppressed to an even lower threshold, which requires a correspondingly higher repetition rate. Specifically, for a single-SDK infidelity of $10^{-4}$, the repetition rate for the ML pulses must exceed 200 MHz to render timing-optimization errors negligible compared to the SDK errors. Achieving a stricter single-SDK infidelity of $10^{-6}$ would push this requirement into the GHz regime, presenting a significant challenge for current experimental setups.

\section{Conclusion}

In summary, we have proposed a programmable pulse scheme based on CW modulation to address two major bottlenecks in trapped-ion ultrafast gates. First, to overcome the high error sensitivity of conventional SDKs, we employ the RAP technique for robust SDK control. Simulation results demonstrate that for a single-pulse duration of $1$~ns, the RAP-based SDK exhibits strong robustness against laser intensity noise in the sub-MHz regime. By designing specific frequency-sweeping profiles, we can further elevate the upper frequency bound of the intensity noise tolerated by the SDK. Second, programmable pulses enable the precise manipulation of pulse timing. The repetition rates of conventional mode-locked lasers are typically limited to tens of MHz, which severely restricts the achievable fidelity in gate optimization. The programmable pulse scheme circumvents this repetition-rate constraint, ensuring that the overall gate fidelity is predominantly limited only by the accumulated errors of individual SDKs.

In our current simulations, the single-SDK infidelity is limited to approximately $10^{-4}$, which is dominated primarily by spontaneous emission. This limitation can be mitigated by further increasing the single-photon detuning. For instance, by utilizing a 355 nm laser, the spontaneous emission error can be suppressed to the $10^{-5}$ level~\cite{campbell2010ultrafast}. However, as the single-photon detuning reaches the scale of tens of THz, coupling to the $^2P_{3/2}$ fine-structure level becomes non-negligible. Overall, our approach provides a highly robust and practical pathway for realizing high-fidelity ultrafast entangling gates, advancing scalable ion-trap quantum computing.

\begin{acknowledgments}
    We thank Sai-Jun Wu for the discussion of pulse laser generation. 
    This work was supported by the National Key Research and Development Program of China (Grant No. 2024YFA1409403), 
    the National Natural Science Foundation of China (Grant No. 11734015, and No. 12204455), Quantum Science and Technology National Science and Technology Major Project (Grant No. 2021ZD0301604 and No. 2021ZD0301200),
    and the Key Research Program of Frontier Sciences, CAS (Grant No. QYZDY-SSWSLH003).
\end{acknowledgments}

\appendix

\section{Effective two-level Hamiltonian with time-dependent two-photon detuning}
\label{app:effective_hamiltonian}
Consider a $\Lambda$-type three-level system interacting with a coherent Raman field. Setting $|0\rangle$ as the energy zero, the total Hamiltonian $H = H_{\mathrm{atom}} + H_{\mathrm{int}}(t)$ under the dipole and rotating-wave approximations is:
\begin{align}
    H_{\mathrm{atom}} &= \hbar\delta_{\mathrm{HF}} |1\rangle\langle 1| + \hbar\omega_e |e\rangle\langle e| \\
    H_{\mathrm{int}}(t) &= \frac{\hbar}{2} \Bigl[ \Omega_1 e^{i(\vec{k}_1\cdot\vec{x} - \omega_1 t)} |e\rangle\langle 0| \nonumber\\
    &\quad + \Omega_2 e^{i\left(\vec{k}_2\cdot\vec{x} - \int_0^t \omega_2(\tau)d\tau\right)} |e\rangle\langle 1| + \text{H.c.} \Bigr]
\end{align}
Define the time-dependent two-photon phase difference $\Delta\phi(t) = \int_0^t [\omega_1 - \omega_2(\tau)] d\tau$. Applying the unitary transformation $U_1(t) = \exp\left( -i \omega_1 t |e\rangle\langle e| \right)$ to eliminate fast-oscillating optical terms, and defining the single-photon detuning $\Delta = \omega_e - \omega_1$, the Hamiltonian in the rotating frame becomes:
\begin{equation}
    \begin{aligned}
        &H_1(t) = \hbar\Delta |e\rangle\langle e| + \hbar\delta_{\mathrm{HF}} |1\rangle\langle 1| \\
        &+ \frac{\hbar}{2} \Bigl[ \Omega_1 e^{i\vec{k}_1\cdot\vec{x}} |e\rangle\langle 0| + \Omega_2 e^{i(\vec{k}_2\cdot\vec{x} + \Delta\phi(t))} |e\rangle\langle 1| + \text{H.c.} \Bigr]
    \end{aligned}
\end{equation}
z
Under the large detuning condition ($\Delta \gg \Omega_1, \Omega_2$), the excited state population is negligible and evolves rapidly. Setting $\dot{c}_{e} \approx 0$ from the Schrödinger equation adiabatically eliminates the excited state:
\begin{equation}
    c_{e} \approx -\frac{1}{2\Delta} \left( \Omega_1 e^{i\vec{k}_1\cdot\vec{x}} c_{0} + \Omega_2 e^{i(\vec{k}_2\cdot\vec{x} + \Delta\phi(t))} c_{1} \right)
\end{equation}

Substituting this back into the Schrödinger equations, absorbing the AC Stark shifts into $\delta_{\mathrm{HF}}$, and introducing the effective two-photon Rabi frequency $\Omega(t) = -\frac{\Omega_1 \Omega_2^*}{2\Delta}$ and the wavevector difference $\Delta\vec{k} = \vec{k}_1 - \vec{k}_2$ with its axial projection $\Delta\vec{k}\cdot\vec{x} = 2\zeta k x$, the system simplifies to an effective two-level Hamiltonian with a time-dependent phase:
\begin{equation}
    H_{\mathrm{eff}}(t) = \frac{\hbar}{2} \left\{ \delta_{\mathrm{HF}} \sigma_z + \left[ \Omega(t) e^{i(2\zeta k x - \Delta\phi(t))} \sigma_+ + \text{H.c.} \right] \right\}
\end{equation}

To eliminate the time-dependent phase $e^{-i\Delta\phi(t)}$ in the off-diagonal terms, we apply a second unitary transformation $U_2(t) = \exp\left( -i \frac{\Delta\phi(t)}{2} \sigma_z \right)$, yielding the gauge potential term:
\begin{equation}
    -i\hbar U_2^\dagger \dot{U}_2 = -\frac{\hbar}{2} \Delta\dot{\phi}(t) \sigma_z = -\frac{\hbar}{2} [\omega_1 - \omega_2(t)] \sigma_z
\end{equation}

The ladder operators transform as $U_2^\dagger \sigma_+ U_2 = \sigma_+ e^{i\Delta\phi(t)}$. Grouping the diagonal terms and defining the time-dependent two-photon detuning $\delta(t) \equiv \delta_{\mathrm{HF}} - [\omega_1 - \omega_2(t)]$, we obtain the final interaction Hamiltonian:
\begin{equation}
    H_{\mathrm{I}} = \frac{\hbar}{2}\left[\Omega(t)\left(e^{i2\zeta kx}\sigma_+ + e^{-i2\zeta kx}\sigma_-\right) + \delta(t)\sigma_z\right],
    \label{eq:H_derivation}
\end{equation}
\section{Adiabatic Evolution Process of the RAP Pulse}
\label{app:evolution_RAP}
\subsection{Instantaneous eigenstates and adiabatic following}
In the rotating frame, the effective two-level Hamiltonian for a single ion is given by
\begin{equation}
    H_{\mathrm{I}}(t) = \frac{\hbar}{2} \begin{pmatrix}
    \delta(t) & \Omega(t)e^{i\phi} \\
    \Omega(t)e^{-i\phi} & -\delta(t)
    \end{pmatrix},
\end{equation}
where $\phi = 2\zeta kx$ is the local spatial phase, and the basis is chosen as $\{|1\rangle, |0\rangle\}$. 
The instantaneous eigenvalues of the system are $E_{\pm}(t) = \pm\frac{\hbar}{2}\Omega_{\mathrm{eff}}(t)$, where $\Omega_{\mathrm{eff}}(t) = \sqrt{\Omega(t)^2 + \delta(t)^2}$. By introducing the mixing angle $\vartheta(t)$ defined by $\tan(\vartheta(t)) = \Omega(t)/\delta(t)$, the corresponding instantaneous eigenstates are
\begin{equation}
    \begin{aligned}
        |+\rangle_t &= \cos\frac{\vartheta(t)}{2}|1\rangle + \sin\frac{\vartheta(t)}{2}e^{-i\phi}|0\rangle, \\
        |-\rangle_t &= \sin\frac{\vartheta(t)}{2}e^{i\phi}|1\rangle - \cos\frac{\vartheta(t)}{2}|0\rangle.
    \end{aligned}
\end{equation}
According to the applied RAP profiles $I(t)$ and $\delta(t)$, the detuning sweeps from $\delta_0 > 0$ to $-\delta_0 < 0$ while the Rabi frequency begins and ends at zero. The mixing angle therefore rotates from $\vartheta(0) = 0$ to $\vartheta(\tau) = \pi$.
Under the adiabatic condition $|\dot{\vartheta}(t)| \ll \Omega_{\mathrm{eff}}(t)$, the system remains in its instantaneous eigenstate. Consequently, an initial ground state $|0\rangle = -|-\rangle_0$ adiabatically follows the $|-\rangle_t$ branch and evolves into $-|-\rangle_\tau = -e^{i\phi}|1\rangle$. Symmetrically, an initial excited state $|1\rangle = |+\rangle_0$ follows the $|+\rangle_t$ branch and ends as $|+\rangle_\tau = e^{-i\phi}|0\rangle$. The spin flip is therefore deterministic and insensitive to the exact pulse area.

\subsection{Derivation of the RAP Evolution Operators}
To construct the unitary evolution operator for the RAP pulse, we solve the time-dependent Schr\"odinger equation $i\hbar \partial_t |\Psi(t)\rangle = H_{\mathrm{I}}(t)|\Psi(t)\rangle$. Under the adiabatic approximation, the state evolves along the instantaneous eigenstates $|n\rangle_t \in \{|+\rangle_t, |-\rangle_t\}$, accumulating both geometric ($\gamma_{\mathrm{g}}^n$) and dynamical ($\gamma_{\mathrm{d}}^n$) phases:
\begin{equation}
    |\Psi(t)\rangle = \sum_{n=\pm} c_n(0) e^{i(\gamma_{\mathrm{g}}^n(t) + \gamma_{\mathrm{d}}^n(t))} |n\rangle_t.
\end{equation}

The geometric phase is determined by the time evolution of the basis vectors:
\begin{equation}
    \gamma_{\mathrm{g}}^\pm = i\int_0^\tau \langle \pm|_t \left( \frac{\partial}{\partial t} |\pm\rangle_t \right) dt.
\end{equation}
Given the instantaneous eigenstates parameterized by the local spatial phase $\phi = 2\zeta kx$ and the mixing angle $\vartheta(t)$:
\begin{equation}
    \begin{aligned}
        |+\rangle_t &= \cos\frac{\vartheta(t)}{2}|1\rangle + \sin\frac{\vartheta(t)}{2}e^{-i\phi}|0\rangle, \\
        |-\rangle_t &= \sin\frac{\vartheta(t)}{2}e^{i\phi}|1\rangle - \cos\frac{\vartheta(t)}{2}|0\rangle.
    \end{aligned}
\end{equation}
Since the spatial phase $\phi$ is constant, the time derivative only acts on $\vartheta(t)$. For the $|+\rangle_t$ branch:
\begin{equation}
    \frac{\partial}{\partial t} |+\rangle_t = -\frac{\dot{\vartheta}}{2}\sin\frac{\vartheta}{2}|1\rangle  + \frac{\dot{\vartheta}}{2}\cos\frac{\vartheta}{2}e^{-i\phi}|0\rangle.
\end{equation}
Evaluating the inner product yields:
\begin{equation}
    \langle +|_t \left( \frac{\partial}{\partial t} |+\rangle_t \right)= 0.
\end{equation}
The same cancellation holds for the $|-\rangle_t$ branch by symmetry. The geometric phase is therefore identically zero:
\begin{equation}
    \gamma_{\mathrm{g}}^\pm = 0.
\end{equation}

The dynamical phase is the time integral of the instantaneous eigenenergies $E_\pm(t) = \pm\frac{\hbar}{2}\sqrt{\Omega(t)^2 + \delta(t)^2}$:
\begin{equation}
    \gamma_{\mathrm{d}}^\pm = -\frac{1}{\hbar}\int_0^\tau E_\pm(t) dt = \mp\Theta(\tau),
\end{equation}
where we define the integrated dynamic phase as $\Theta(\tau) = \frac{1}{2}\int_0^\tau \sqrt{\Omega(t)^2 + \delta(t)^2} dt$.

Applying these phases to the adiabatic mappings from the initial state ($\vartheta=0$) to the final state ($\vartheta=\pi$), the computational basis states transform as:
\begin{equation}
    \begin{aligned}
        |1\rangle &\rightarrow e^{i\gamma_{\mathrm{d}}^+}|+\rangle_\tau = e^{-i\Theta(\tau)}e^{-i\phi}|0\rangle, \\
        |0\rangle &\rightarrow -e^{i\gamma_{\mathrm{d}}^-}|-\rangle_\tau = -e^{i\Theta(\tau)}e^{i\phi}|1\rangle.
    \end{aligned}
\end{equation}

Writing this transformation in terms of the Pauli ladder operators ($\sigma_+ = |1\rangle\langle 0|$ and $\sigma_- = |0\rangle\langle 1|$), we obtain the unitary operator for a single ion:
\begin{equation}
    U^-(\zeta) = -e^{i(2\zeta kx+\Theta(\tau))}\sigma_+ + e^{-i(2\zeta kx+\Theta(\tau))}\sigma_-.
\end{equation}

\section{Derivation of the Total Fast Gate Evolution}
\label{app:total_gate}

To derive the total evolution operator $U_{\mathrm{total}}$ for the fast gate, we trace the effect of $N$ SDK pulse groups applied to a two-ion string. Defining the collective spin operators $\Sigma_{\mathrm{c}} = \sigma_z^1 + \sigma_z^2$ and $\Sigma_{\mathrm{s}} = \sigma_z^1 - \sigma_z^2$, the instantaneous kick operator at time $t_k$ with amplitude $z_k = -n_k\zeta$ is:
\begin{equation}
    U_k = \hat{D}_{\mathrm{c}}\left(2i z_k \eta_{\mathrm{c}} \Sigma_{\mathrm{c}}\right) \hat{D}_{\mathrm{s}}\left(2i z_k \eta_{\mathrm{s}} \Sigma_{\mathrm{s}}\right).
\end{equation}

In the interaction picture defined by the free motional Hamiltonian $H_{\mathrm{free}} = \hbar\omega_{\mathrm{c}} a_{\mathrm{c}}^\dagger a_{\mathrm{c}} + \hbar\omega_{\mathrm{s}} a_{\mathrm{s}}^\dagger a_{\mathrm{s}}$, the displacement operators acquire a time-dependent phase via $a_m(t) = a_m e^{-i\omega_m t}$. The interaction-picture kick operator becomes:
\begin{equation}
    \tilde{U}_k(t_k) = \hat{D}_{\mathrm{c}}\left(2i z_k \eta_{\mathrm{c}} \Sigma_{\mathrm{c}} e^{i\omega_{\mathrm{c}} t_k}\right) 
     \times \hat{D}_{\mathrm{s}}\left(2i z_k \eta_{\mathrm{s}} \Sigma_{\mathrm{s}} e^{i\omega_{\mathrm{s}} t_k}\right).
\end{equation}

The total evolution operator $U_{\mathrm{total}}$ is the time-ordered product of these operators:
\begin{equation}
    U_{\mathrm{total}} = \prod_{k=1}^N \tilde{U}_k(t_k).
\end{equation}
Since the COM and stretch modes commute, we evaluate their products independently using the Baker-Campbell-Hausdorff (BCH) formula $\hat{D}(\alpha)\hat{D}(\beta) = \hat{D}(\alpha+\beta)\exp[i\operatorname{Im}(\alpha\beta^*)]$. Applying this iteratively yields:
\begin{equation}
    \begin{aligned}
        &\prod_{k=1}^N \hat{D}_m\left(\Delta \alpha_m^{(k)} \Sigma_m\right) \\
        &= \hat{D}_m\left(\Sigma_m \sum_{k=1}^N \Delta \alpha_m^{(k)}\right) \\
        &\quad \times \exp\left[ i \Sigma_m^2 \sum_{j>k}^N \operatorname{Im}\left(\Delta \alpha_m^{(j)} (\Delta \alpha_m^{(k)})^*\right) \right].
    \end{aligned}
\end{equation}

By defining the total motional displacement amplitude $\alpha_m$ and the accumulated geometric phase angle $\theta_m$ for each mode $m \in \{c, s\}$ as:
\begin{equation}
    \begin{aligned}
        \alpha_m &= 2i\eta_m \sum_{k=1}^N z_k e^{i\omega_m t_k}, \\
        \theta_m &= 4\eta_m^2 \sum_{j=2}^N \sum_{k=1}^{j-1} z_j z_k \sin[\omega_m(t_j - t_k)],
    \end{aligned}
\end{equation}
the total evolution simplifies to:
\begin{equation}
    U_{\mathrm{total}} = \hat{D}_{\mathrm{c}}(\alpha_{\mathrm{c}}\Sigma_{\mathrm{c}})\hat{D}_{\mathrm{s}}(\alpha_{\mathrm{s}}\Sigma_{\mathrm{s}})\times \exp\left[i \left( \theta_{\mathrm{c}} \Sigma_{\mathrm{c}}^2 + \theta_{\mathrm{s}} \Sigma_{\mathrm{s}}^2 \right)\right].
\end{equation}

Finally, we expand the collective spin operators squared:
\begin{equation}
    \begin{aligned}
        \Sigma_{\mathrm{c}}^2 &= (\sigma_z^1 + \sigma_z^2)^2 = 2 + 2\sigma_z^1\sigma_z^2, \\
        \Sigma_{\mathrm{s}}^2 &= (\sigma_z^1 - \sigma_z^2)^2 = 2 - 2\sigma_z^1\sigma_z^2.
    \end{aligned}
\end{equation}
Substituting these into the exponent gives the total accumulated phase:
\begin{equation}
    i \left[ \theta_{\mathrm{c}} \Sigma_{\mathrm{c}}^2 + \theta_{\mathrm{s}} \Sigma_{\mathrm{s}}^2 \right]= i2(\theta_{\mathrm{c}} + \theta_{\mathrm{s}}) - i2(\theta_{\mathrm{s}} - \theta_{\mathrm{c}})\sigma_z^1\sigma_z^2.
\end{equation}
The first term $i2(\theta_{\mathrm{c}} + \theta_{\mathrm{s}})$ acts as a global phase and can be omitted. By defining the two-qubit entangling phase as $\Phi = 2(\theta_{\mathrm{s}} - \theta_{\mathrm{c}})$, we obtain the final form of the fast gate evolution operator:
\begin{equation}
    U_{\mathrm{total}} = \hat{D}_{\mathrm{c}}(\alpha_{\mathrm{c}}\Sigma_{\mathrm{c}})\hat{D}_{\mathrm{s}}(\alpha_{\mathrm{s}}\Sigma_{\mathrm{s}}) \exp(-i\Phi\sigma_z^1\sigma_z^2).
\end{equation}

\nocite{*}

\bibliography{apssamp}
\end{document}